\documentclass[aps,prb,twocolumn,superscriptaddress]{revtex4}
\usepackage{graphicx}
\usepackage{amssymb}
\usepackage{amsmath}
\usepackage{graphicx}
\usepackage{epsfig}
\usepackage{color}
\usepackage{bm}

\begin{document}

\title{Dynamic charge Kondo effect and a slave fermion approach to the Mott transition}
\author{Zhuoqing Long}
\affiliation{Beijing National Laboratory for Condensed Matter Physics, Institute of Physics,
Chinese Academy of Science, Beijing 100190, China}
\affiliation{School of Physical Sciences, University of Chinese Academy of Sciences, Beijing 100190, China}
\author{Jiangfan Wang}
\affiliation{Beijing National Laboratory for Condensed Matter Physics, Institute of Physics,
	Chinese Academy of Science, Beijing 100190, China}
\author{Yi-feng Yang}
\email[]{yifeng@iphy.ac.cn}
\affiliation{Beijing National Laboratory for Condensed Matter Physics,  Institute of Physics, 
Chinese Academy of Science, Beijing 100190, China}
\affiliation{School of Physical Sciences, University of Chinese Academy of Sciences, Beijing 100190, China}
\affiliation{Songshan Lake Materials Laboratory, Dongguan, Guangdong 523808, China}
\date{\today}

\begin{abstract}
Mott transition plays a key role in strongly correlated physics but its nature is not yet fully understood. Motivated by recent development of Schwinger boson approach for the Kondo lattice, we propose in this work a novel slave fermion algorithm to study the Mott transition. Upon local approximation, our method yields a phase diagram with a zero-temperature continuous (Mott) metal-insulator transition at finite Coulomb interaction $U$ for the half-filled one-band Hubbard model on a square lattice, and the resistivity exhibits a critical scaling around the quantum Widom line. We argue that the Mott transition may be associated with a dynamic charge Kondo effect of local degenerate doublon and holon states, causing sharp resonances on the doublon/holon and electron spectra. The transition is pushed to $U=0$ once intersite antiferromagnetic correlations are included, in agreement with exact numerical calculations. Our approach captures some essential features of the Mott transition and offers an alternative angle to view this important problem. It can be extended to study other correlated electron models with more complicated local interactions.
\end{abstract}

\maketitle

\section{Introduction}
The Mott transition is a metal-insulator transition driven by local Coulomb interactions \cite{Hubbard1963,Brinkman-Rice1970}. Its nature remains obscure for over half a century \cite{MetalInsulatorTransitionsReview1998} due to the lack of an obvious order parameter \cite{Dobrosavljevic2012Book, Phillips2022Z2symmetry}, which makes it a key problem to explore new physics beyond the Landau paradigm of phase transitions. Experimentally, it has been observed in a wide spectrum of materials such as transition metal oxides \cite{McWhan1973V2O3PhaseDigram,Limelette2003V2O3}, organic compounds \cite{Lefebvre2000kappaCl,Kagawa2005kappaCl_Scaling, Kurosaki2005kappaCu2(CN)3,Furukawa2018kappaCu2(CN)3}, and transition-metal dichalcogenides (TMDs) \cite{Ghiotto2021TMD,Li2021TMD}. There are two major features in the electron spectra associated with the Mott transition: the high-energy lower and upper Hubbard bands and the low-energy quasiparticle peak which is suppressed as the system turns from a metal to an insulator \cite{Mo2003V2O3PES, Xu2014NiS2-xSexARPES}. Near the transition, a quantum critical scaling has also been reported in resistivity \cite{Furukawa2015MottCriticality}. Theoretically, the Mott transition has been extensively explored using various analytical  and numerical methods \cite{Arovas2022HubbardReview,Qin2022NumericalReview}, including studies based on parton theories \cite{Kotliar1986KRSlaveBoson, Li1989SRIKR, Fresard1992SRIKR, CupratesReview2006, Kaga1992SlaveBosonCompare, Florens2002SlaveRotor, Florens2004SlaveRotor, Lee2005SpinLiquid, 
Kim2006SU(2)SlaveRotor, Hermele2007SU(2)SlaveRotor, Zhao2007SlaveRotorClusterMFT, Senthil2008CriticalFS, Senthil2008SlaveRotor, 
Podolsky2009SlaveRotor3DCase, Xu2022TMDMottTransition, de'Medici2005Z2SlaveSpin1, de'Medici2010Z2SlaveSpin1, Ruegg2010Z2SlaveSpin2, Nandkishore2010OrthogonalMetal, Zitko2015MottZ2, Yu2012U(1)SlaveSpin, Yu2017U(1)SlaveSpin}, doublon-holon binding scenario \cite{Castellani1979DHBinding, Kaplan1982DHBinding, Capello2005DHBinding, Yokoyama2006DHBinding, MottReview2010, Zhou2014DHBinding, Sato2014DHBinding, 	Prelovsek2015DHBinding, Han2016SlaveFermion, Han2019SlaveFermion, Zhou2020DHBinding}, charge-$2e$ boson theory \cite{Leigh2007Charge2eBoson,Leigh2009Charge2eBoson}, Hatsugai-Kohmoto (HK) model \cite{HKmodel1992, Phillips2020superconductivity,Phillips2022Z2symmetry}, dynamical mean-field theory (DMFT) and its extensions \cite{DMFTReview1996,clusterDMFTReview2005, CTQMCReview2011, diagrammaticReview2018}, comparison of different numerical approaches \cite{LeBlanc2015Benchmark, Schafer2021Benchmark}, and many others \cite{Meinders1993DSWT, Eskes1994DSWT, Dzyaloshinskii2003LuttingerTheorem,Borejsza2004NLSM, Held2000PAMandHubbard, Hu2020ClassicalCorrespondence, Hu2020PAMandHubbard, 
Byczuk2012RelativeEntropy, Walsh2019Information, Sen2020Topological}.
Among them, DMFT can yield both features as well as the resistivity scaling \cite{Terletska2011MottCriticality, Vucicevic2013MottCriticality, Vucicevic2015MottCriticality, Lee2016MottCriticalityAndSpinon, Dasari2017MottSusceptibilityScaling, Eisenlohr2019MottCriticality} but predicted a first-order Mott transition \cite{Kotliar1999DMFTLandauTheory, Kotliar2000DMFTLandauTheory} at finite Coulomb interaction for the half-filled one-band Hubbard model. On the other hand, exact determinant quantum Monte Carlo (DQMC) simulation on a square lattice predicted a continuous transition at $U=0$ \cite{Hirsch1985DQMC, White1989DQMC, Preuss1995FourPeakQMC, Schafer2015AFM_and_Mott, Vitali2016U=0gap, Simkovic2020U=0gap}, possibly due to intersite antiferromagnetic (AFM) correlations \cite{Schafer2015AFM_and_Mott}. Since many effects are involved in these calculations, it is difficult to clarify the role of each factor and establish a clear picture of the Mott transition. 

Slave particle approaches have also been applied to understand the Mott transition, which describe the electrons by fractionalized spin and charge degrees of freedom such as spinons, holons, doublons, or rotors \cite{Kotliar1986KRSlaveBoson, Li1989SRIKR, Fresard1992SRIKR, CupratesReview2006, Kaga1992SlaveBosonCompare, Florens2002SlaveRotor, Florens2004SlaveRotor, Lee2005SpinLiquid, Kim2006SU(2)SlaveRotor, Hermele2007SU(2)SlaveRotor, Zhao2007SlaveRotorClusterMFT, Senthil2008CriticalFS, Senthil2008SlaveRotor, Podolsky2009SlaveRotor3DCase, Xu2022TMDMottTransition}. 
These approaches are less rigorous compared to exact numerical methods, but may provide more intuitive pictures of the transition. To date, most of these works adopted a fermionic representation of the spinons and treated the charge degrees of freedom as bosons. As a result, the Mott transition may be naturally characterized as boson condensation. By contrast, fermionic representation \cite{Yoshioka1989SlaveFermion,Jayaprakash1989SlaveFermion,Kane1990SlaveFermion,Chakraborty1990SlaveFermion,Auerbach1991SlaveFermion,Li1992SlaveFermion,Igarashi1992SlaveFermion,Khaliullin1993SlaveFermion} is seldom used for the charge degrees of freedom, although doublon and holon excitations created by $c_{i\sigma}^\dagger n_{i,-\sigma}$ and $c_{i\sigma}(1-n_{i,-\sigma})$ have a  fermion-like form \cite{MottReview2010}. Here $c_{i\sigma}$ and $c_{i\sigma}^\dagger$ are the annihilation and creation operators of electrons and $n_{i\sigma}$ is its density operator. In fact, a fermionic representation with self-consistent Born approximation (SCBA) has recently been applied to analyze the doublon-holon binding and found good agreement with exact DQMC simulations \cite{Han2016SlaveFermion,Han2019SlaveFermion}. 

In this work, we propose an alternative approach based on the fermionic representation. Our method is motivated by the recent development of Schwinger boson approach \cite{Parcollet1997SchwingerBoson,Coleman2005SchwingerBoson,Rech2006SchwingerBoson,Komijani2018SchwingerBoson,Komijani2019SchwingerBoson,Wang2020SchwingerBoson,Wang2021SchwingerBoson,Han2021SchwingerBoson,Wang2022SchwingerBosonSpinCurrent,Wang2022SchwingerBosonFMQCP,Wang2022SchwingerBosonMetallicSpinLiquid} to heavy fermion systems. Instead of using Born approximation, we introduce a fermionic auxiliary field to decouple the kinetic term and use self-consistent one-loop approximation to calculate the quasiparticle spectra. As an example, we apply this method to the one-band Hubbard model at half filling and derive successfully a $T$-$U$ phase diagram with a continuous metal-insulator transition at zero temperature. Near the transition, the resistivity exhibits a critical scaling around the quantum Widom line \cite{Vucicevic2013MottCriticality}. We find that the transition may be understood as a dynamic charge Kondo effect of the local degenerate doublon and holon states, with bosonic spinons playing the role of fluctuating hybridization fields. Comparison with DMFT indicates that the DMFT first-order transition may originate from vertex corrections beyond one-loop approximation. Including intersite AFM correlations drives the transition to $U=0$, in good agreement with DQMC prediction. Hence, our approach captures the essential aspects of the Mott transition and is a useful tool to explore the roles of different ingredients during the transition. It can be easily extended to other more complicated models and offer novel perspective on their correlated electron physics.

\section{Method}
\begin{figure}
	\includegraphics[width=8.6cm]{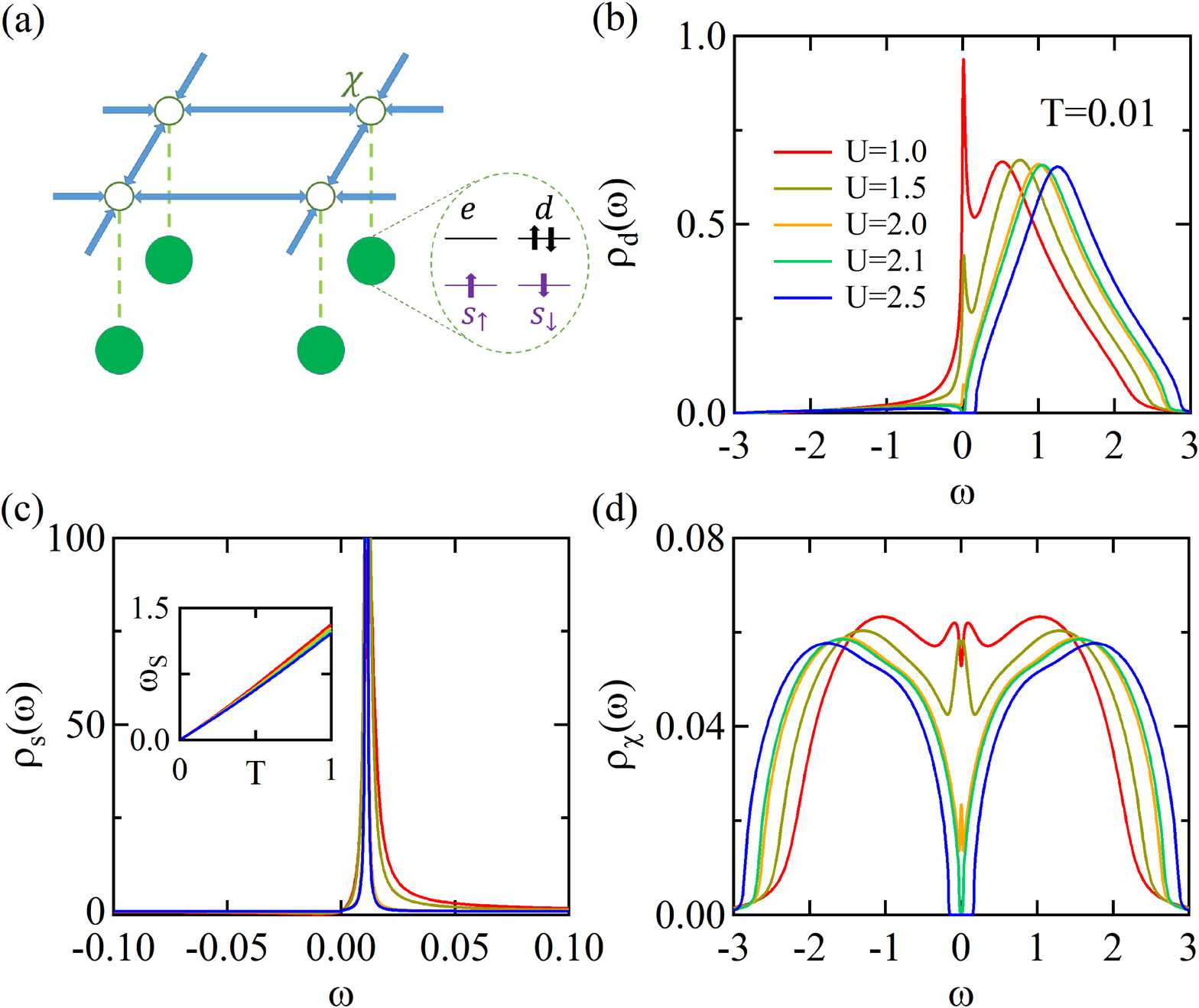}
	\caption{(a) Illustration of the effective model with the fermionic auxiliary field $\chi$ and the local slave particles. (b)-(d) DOS of doublon, spinon, and auxiliary $\chi$ field for different values of $U$ at $T=0.01$. The inset of (c) shows the roughly linear relationship between the spinon peak position $\omega_s$ and the temperature.}
	\label{fig1}
\end{figure}

We start with the half-filled one-band Hubbard model on a square lattice:
\begin{equation}
H=-\sum_{ij\sigma}t_{ij}c_{i\sigma}^\dagger c_{j\sigma}+U\sum_{i}\left( n_{i\uparrow}-\frac12\right) \left( n_{i\downarrow}-\frac12\right),
\label{Hubbard model}
\end{equation}
which contains a kinetic term and a Hubbard $U$-term. In the slave fermion representation, the physical electron operator is rewritten as $c_{i\sigma}=e^\dagger_is_{i\sigma}+\sigma s^\dagger_{i,-\sigma}d_i$, where $d_i$ and $e_i$ are fermionic operators denoting the doubly occupied (doublon) and empty (holon) states, respectively, and $s_{i\sigma}$ describe bosonic spinons for the singly occupied states. The physical Hilbert space is constrained by requiring $Q_i\equiv e^\dagger_ie_i+d^\dagger_id_i+\sum_\sigma s^\dagger_{i\sigma}s_{i\sigma}=1$. In this way, the Hubbard-$U$ term becomes quadratic, but the kinetic term turns into a more complicated quartic form, which may be decoupled via a Hubbard-Stratonovich transformation by introducing a fermionic auxiliary field $\chi_{i\sigma}$ \cite{Pairault1998StrongCoupling, Pairault2000StrongCoupling,Rubtsov2008DualFermion,Brener2008DualFermion,Rubtsov2009DualFermion,Li2015DualFermion}. The final effective action reads
\begin{eqnarray}
\mathcal{L}&=&\mathcal{L}_0+\frac{U}{2}\sum_i(\bar d_id_i+\bar e_i e_i)+\sum_i\lambda_i(Q_i-1)\\
&-&\sum_{ij\sigma}\mathcal G_{ij}^{-1}\bar\chi_{i\sigma}\chi_{j\sigma}+\sum_{i\sigma}[\bar\chi_{i\sigma}(\bar e_{i}s_{i\sigma}+\sigma \bar s_{i,-\sigma}d_{i})+\text{H.c.}],\notag
\label{interaction term}
\end{eqnarray}
where $\mathcal{L}_0=\sum_i\left(\bar d_i \partial_\tau d_i+\bar e_i\partial_\tau e_i+\sum_\sigma\bar s_{i\sigma}\partial_\tau s_{i\sigma}\right)$, $\lambda_i$ is the Lagrange multiplier, and $\mathcal G_{ij}^{-1}=-(t^{-1})_{ij}$ with $t$ being the hopping matrix with site indices. As illustrated in Fig. \ref{fig1}(a), only the auxiliary $\chi$ field carries spatial correlation information in this model, while the local slave particles $d_i$, $e_i$, and $s_{i\sigma}$ can only hop on the lattice by coupling to the $\chi$ field through a three-particle vertex. Such a form actually covers a large family of strongly correlated models, similar to that in the dual fermion approach \cite{Rubtsov2008DualFermion,Brener2008DualFermion,Rubtsov2009DualFermion,Li2015DualFermion}. The above procedure can be readily extended to other models with more complex local interactions. By requiring $\mathcal G^{-1}(\bm k,\text{i}\omega_n)=(\text{i}\omega_n-\varepsilon_{\bm k})$, we may also obtain the periodic Anderson model with local degrees of freedom represented by slave particles.

There are different ways to solve the above action. We may use DMFT by approximating $\mathcal G$ with a self-consistent local Weiss field \cite{DMFTReview1996}. But DMFT is a black box incorporating many effects that are hard to extract. Here, motivated by the recent development of Schwinger boson approach for studying heavy fermion quantum phase transitions \cite{Komijani2018SchwingerBoson,Komijani2019SchwingerBoson,Wang2020SchwingerBoson,Wang2021SchwingerBoson,Han2021SchwingerBoson,Wang2022SchwingerBosonSpinCurrent,Wang2022SchwingerBosonFMQCP,Wang2022SchwingerBosonMetallicSpinLiquid}, we adopt one-loop approximation and solve the following self-energy equations:
\begin{eqnarray}
&&\Sigma_{\chi}(\text{i}\omega_n)=\frac{1}{\beta}\sum_mG_s(\text{i}\nu_m)[G_e(\text{i}\omega_{m-n})-G_d(\text{i}\omega_{m+n})],\notag\\
&&\Sigma_s(\text{i}\nu_m)=\dfrac{1}{\beta }\sum_{n}G_\chi(\text{i}\omega_n)\left[G_d(\text{i}\omega_{n+m})-G_e(\text{i}\omega_{m-n})  \right],\notag\\
&&\Sigma_{d/e}(\text{i}\omega_n)=-\frac{2}{\beta}\sum_mG_s(\text{i}\nu_m)G_\chi(\text{i}\omega_{n-m}),
\label{eq:SelfE}
\end{eqnarray}
where $\text{i}\omega$ ($\text{i}\nu$) are the fermionic (bosonic) Matsubara frequencies, $\Sigma$ are the local self-energies of all slave particles and auxiliary fields, and $G$ are their full local Green's functions to be determined self-consistently. The holon and doublon self-energies are the same due to the particle-hole symmetry. For simplicity, we have adopted a local approximation and ignore the moment dependency in the self-energies. The Lagrange multipliers are also approximated by their mean-field value, $\lambda_i=\lambda$. The one-loop approximation is numerically more efficient than DMFT. In this form, we may have a more intuitive picture on physics, in the cost of numerical exactness. Below we will focus on the square lattice model with only nearest-neighbor hopping $t$, and set the half bandwidth $D$ of the free electrons to unity so that $t=1/4$. Note that all calculations were performed in real frequency.

\section{Results and Discussions}
Figures \ref{fig1}(b)-\ref{fig1}(d) plot the resulting density of states (DOS) for the slave particles and the auxiliary field $\chi$. As shown in Fig. \ref{fig1}(b), the doublon (holon) spectra contain a broad peak around $\omega=U/2+\lambda$, which is the bare energy of the local doublon (holon) state. The spectra are, however, greatly broadened due to the coupling with the spinon and $\chi$ fields. The slight spectral weight below the Fermi energy contributes a small doublon occupation number. For large $U$ above $U_c\approx2.1$, we see a finite excitation gap, which gradually diminishes with decreasing $U$ as the whole spectra move leftwards and touch the Fermi energy. For $U<U_c$, a sharp resonance peak appears around $\omega\sim \eta T$ ($\eta\approx 1-2$) for $U<U_c$. Hence $U_c$ marks the Mott metal-insulator transition point below which the charge excitation gap closes. The sharp resonance on the doublon (holon) spectra reflects the emergent electron quasiparticles in the metallic state.

To clarify its origin, we note that a similar peak also appears in the spinon spectra as shown in Fig. \ref{fig1}(c). Its location scales linearly with the temperature, $\omega_s\approx1.1T$ (inset). This may be easily understood from the constraint $\langle Q_i\rangle=1$. Because both doublon and holon occupations are small, the spinon energy must satisfy $n_B(\omega_s)\approx 0.5$ if we ignore its broadening. Here $n_B(\omega)$ is the Bose-Einstein distribution function. We immediately find $\omega_s\approx1.1T$. Hence, the resonance peak in the doublon (holon) spectra comes from the coupling with spinons and the auxiliary $\chi$ field through the three-particle vertex. Since the spinon is a bosonic field, we may also compare it to the Kondo effect \cite{HewsonBook,ColemanBook}, where the Kondo resonance appears when two degenerate local (spin) states couple to conduction electrons through a finite hybridization field. This suggests a possible Kondo-like mechanism in the Hubbard model, albeit for the degenerate doulon and holon states. 

In fact, if we integrate out the bosonic spinons and define $f_i\equiv(d_i, e_i)^{\text T}$ and $\psi_i\equiv (\chi_{i\downarrow}, \bar\chi_{i\uparrow})^{\text T}$, the vertex term in the effective action Eq. (\ref{interaction term}) becomes $g^s_{i,\tau_1-\tau_2}\left[(\bar f_{i\tau_1}\psi_{i\tau_1})(\bar \psi_{i\tau_2}f_{i\tau_2})-(\bar f_{i\tau_1}\text{i}\sigma^y\bar\psi_{i\tau_1})(\psi_{i\tau_2}\text{i}\sigma^y f_{i\tau_2})\right]$, where $g^s_{i\tau}$ is the bare spinon Green's function in imaginary time. This looks quite complicated. However, at $\tau_1=\tau_2$, it reduces to a familiar form, $-4g^s_i\bm S^f_{i}\cdot\bm S^\psi_{i}$, where $\bm S^f_{i}=\bar f_{i}\frac{\bm\sigma}{2}f_{i}$ and $\bm S^\psi_{i}=\bar \psi_{i}\frac{\bm\sigma}{2}\psi_{i}$. In deriving this, we have used the equality, $\bm\sigma_{\alpha\beta}\cdot\bm\sigma_{\mu\nu}=\delta_{\alpha\nu}\delta_{\beta\mu}-(\text{i}\sigma^y)_{\alpha\mu}(\text{i}\sigma^y)_{\beta\nu}$. It can be shown that  $g^s_i=n_B(-\lambda)$, which is always negative for $\lambda>0$, indicating an effective antiferromagnetic coupling. Hence, the three-particle vertex contains some Kondo-like physics, where the doublon/holon are local degenerate states, the $\chi$ field plays the role of conduction electrons, and the spinon contributes the hybridization. This analogy underlies the Mott transition at $U_c$ and the emergence of electron quasiparticles for $U<U_c$.

\begin{figure}
\includegraphics[width=8.6cm]{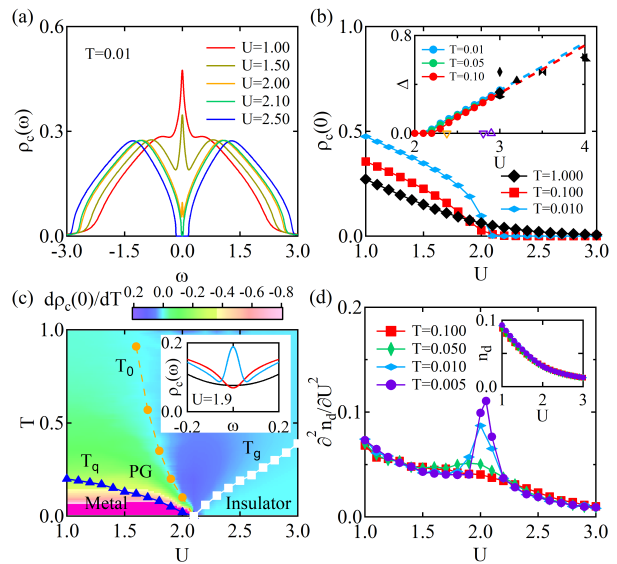}
\caption{(a) The calculated DOS of the physical $c$-electrons at $T=0.01$ for different values of $U$.  (b) DOS at the Fermi energy ($\omega=0$) as a function of $U$ for $T=1.0$, 0.1 and 0.01. The inset shows our calculated gap size $\Delta$ (circle) at different temperatures compared to those of DMFT using different impurity solvers for $T$ varying between 0 and 0.10 (black solid markers) \cite{
Okamoto2005DMFT_SquareLattice,Kusunose2006DMFT_SquareLattice,Fuhrmann2007DMFT_SquareLattice,Zhang2007DMFT_SquareLattice,Wang2009DMFT_SquareLattice,Zitko2009DMFT_SquareLattice,Ribic2018DMFT_SquareLattice}. Also shown are the boundaries of the first-order Mott transition predicted by DMFT at $T=0$ (empty markers: $U_{c1}$: orange; $U_{c2}$: purple) for the square lattice Hubbard model \cite{Zitko2009DMFT_SquareLattice,Park2008CDMFT}.  (c) Intensity plot of $\text{d}\rho_c(0)/\text{d}T$ showing three temperature scales and the metallic, pseudogap (PG), and insulating regions. $T_q$ sets the boundary for quasiparticle resonance, $T_0$ is given by $\text{d}\rho_c(0)/\text{d}T=0$, and $T_g$ is determined by $\rho_c(0)<0.001$ for $U>U_c$. The inset shows the DOS features of three temperatures near the Fermi energy for $U=1.9$. (d) Second derivative $\partial^2n_d/\partial U^2$ as a function of $U$ for different temperatures, showing an increasingly sharp peak at $U_c$ with lowering temperature. For comparison, the inset shows the continuous variation of  $n_d$ itself.}
\label{fig2}
\end{figure}

However, there are two important differences from the usual Kondo effect. First, it happens in the charge degrees of freedom, the effective ``spin" can have a time dependence for $\tau_1\neq\tau_2$, and the bosonic ``hybridization" (spinon) field has a peak at finite frequency. Hence, we may better call it a dynamic charge Kondo effect. Second, most of the holon and doublon spectra are above the Fermi energy, so these states are only slightly occupied. As a result, the Kondo-like resonance only appears when their spectra are close to the Fermi energy, and the interaction-driven Mott transition must be associated with some singular variation in the doublon (holon) occupation. We will see that this is indeed the case. 

For completeness, we also show in Fig. \ref{fig1}(d) the spectra of the auxiliary $\chi$ field. Similarly, it contains two broad peaks at high energies representing the processes of creating a doublon and annihilating a holon, respectively. Around the Fermi energy, the spectra are gapped for $U>U_c$ but develop a finite weight for $U<U_c$. The gap for $U>U_c$ separates the peaks at positive and negative energies and implies that the doublon and holon processes are independent and do not cooperate to form electron quasiparticles. For $U<U_c$, the finite spectra at $\omega=0$ imply that the $\chi$ field becomes itinerant, through which all low-energy slave particle excitations become movable on the lattice. In addition, the gapped doublon/holon and $\chi$ spectra also prohibit spinon decay, so the spinon peak at $\omega_s$ is sharp for large $U$ but damped for small $U$, as is seen in Fig. \ref{fig1}(c).

As in the usual spin Kondo effect, the dynamic charge Kondo effect for $U<U_c$ leads to a superposition of the holon and doublon states through coupling to the $\chi$ field. Such a superposition is necessary for the emergence of electron-like quasiparticles because $c_{i\sigma}=e^\dagger_is_{i\sigma}+\sigma s^\dagger_{i,-\sigma}d_i$ involves both holons and doublons. To see this, we calculate the spectra of physical electrons using the self-energy of the $\chi$ field, $G_c(\bm k,\text{i}\omega_n)=\Sigma_\chi(\text{i}\omega_n)/[1-\epsilon_{\bm k}\Sigma_\chi(\text{i}\omega_n)]$, where $\epsilon_{\bm k}$ is the bare dispersion of electrons. This formula can be derived by making derivative of the action with respect to the hopping parameter $t_{ij}$ \cite{Rubtsov2008DualFermion,Brener2008DualFermion,Rubtsov2009DualFermion,Li2015DualFermion}. The resulting local DOS $\rho_c(\omega)$ is plotted in Fig. \ref{fig2}(a). We find the same features as in the slave particles with two broad peaks near $\pm(U/2+\lambda)$, which can be identified as the upper and lower Hubbard bands. For $U>U_c$, the spectra are also gapped, while for $U<U_c$, we find a sharp quasiparticle peak at $\omega=0$. Obviously, all these features are related to those on the doublon and holon spectra. To see the Mott transition clearly, we plot the electron DOS at zero energy $\rho_c(0)$ in Fig. \ref{fig2}(b). For all temperatures, $\rho_c(0)$ decreases as $U$ grows. At low temperatures, $\rho_c(0)$ drops rapidly to nearly zero at $U_c$, signaling the occurrence of a Mott transition. Our results are supported by the comparison of the Mott gap above $U_c$ with those derived by DMFT using different impurity solvers \cite{
Okamoto2005DMFT_SquareLattice,Kusunose2006DMFT_SquareLattice,Fuhrmann2007DMFT_SquareLattice,Zhang2007DMFT_SquareLattice,Wang2009DMFT_SquareLattice,Zitko2009DMFT_SquareLattice,Ribic2018DMFT_SquareLattice}. As plotted in the inset of Fig. \ref{fig2}(b), the overall agreement is quite good, except that DMFT predicts a first-order transition with a slightly larger lower boundary $U_{c1}\approx 2.3$ at $T=0$ from its insulating solution \cite{Zitko2009DMFT_SquareLattice,Park2008CDMFT}.

To have an overall picture of the transition, we construct in Fig. \ref{fig2}(c) a tentative phase diagram based on the intensity plot of $\text{d}\rho_c(0)/\text{d}T$. We find three temperature scales separating the phase diagram into four regions. For $U>U_c$, the gap in $\rho_c(\omega)$ is filled in gradually by thermal excitations, so $\rho_c(0)$ always increases with increasing temperature ($\text{d}\rho_c(0)/\text{d}T>0$). At low temperature, the thermal effects are suppressed, so we may roughly estimate a crossover temperature $T_g$ by requiring $\rho_c(0)<0.001$ as the upper boundary of the Mott insulator. For $U<U_c$, as shown in the inset of Fig. \ref{fig2}(c), the electron DOS exhibits a quasiparticle peak at very low temperature. Increasing temperature first destroys the quasiparticle peak, causing a local minimum (dip) at the Fermi energy above a certain temperature $T_q$, but $\rho_c(0)$ keeps decreasing till a higher temperature $T_0>T_q$. Above $T_0$, thermal smearing effect takes over and gives a positive sign for $\text{d}\rho_c(0)/\text{d}T$. The region between $T_q$ and $T_0$ is then recognized as a pseudogap region, with a dip in $\rho_c(\omega)$ around the Fermi energy (see the inset for $T=0.1$). All these three temperature scales, $T_q$, $T_g$, and $T_0$, drop continuously to zero as $U$ approaches $U_c\approx2.1$, indicating a zero-temperature Mott transition instead of a finite temperature first-order transition  predicted by DMFT. To determine the order of the transition, we further calculate the double occupancy $n_d\equiv\sum_{i}\langle n_{i\uparrow}n_{i\downarrow}\rangle/\mathcal N=\sum_{i}\langle d^\dagger_{i}d_{i}\rangle/\mathcal N$, where $\mathcal N$ is the number of lattice sites. $n_d$ is therefore also the doublon occupation. The results are shown in Fig. \ref{fig2}(d). We see it also varies continuously with $U$ for all temperatures (inset), but its second derivative $\partial^2n_d/\partial U^2$ exhibits a sharp peak near $U_c$ as $T$ decreases, implying a jump in $\partial n_d/\partial U$ or a slope change in $n_d(U)$ at $U_c$ as $T\rightarrow 0$. Since $n_d$ is proportional to the first derivative of the free energy $F$ with respect to $U$, our method successfully captures the Mott transition and identifies it as a continuous (second-order) quantum phase transition at zero temperature.

\begin{figure}
	\includegraphics[width=8.6cm]{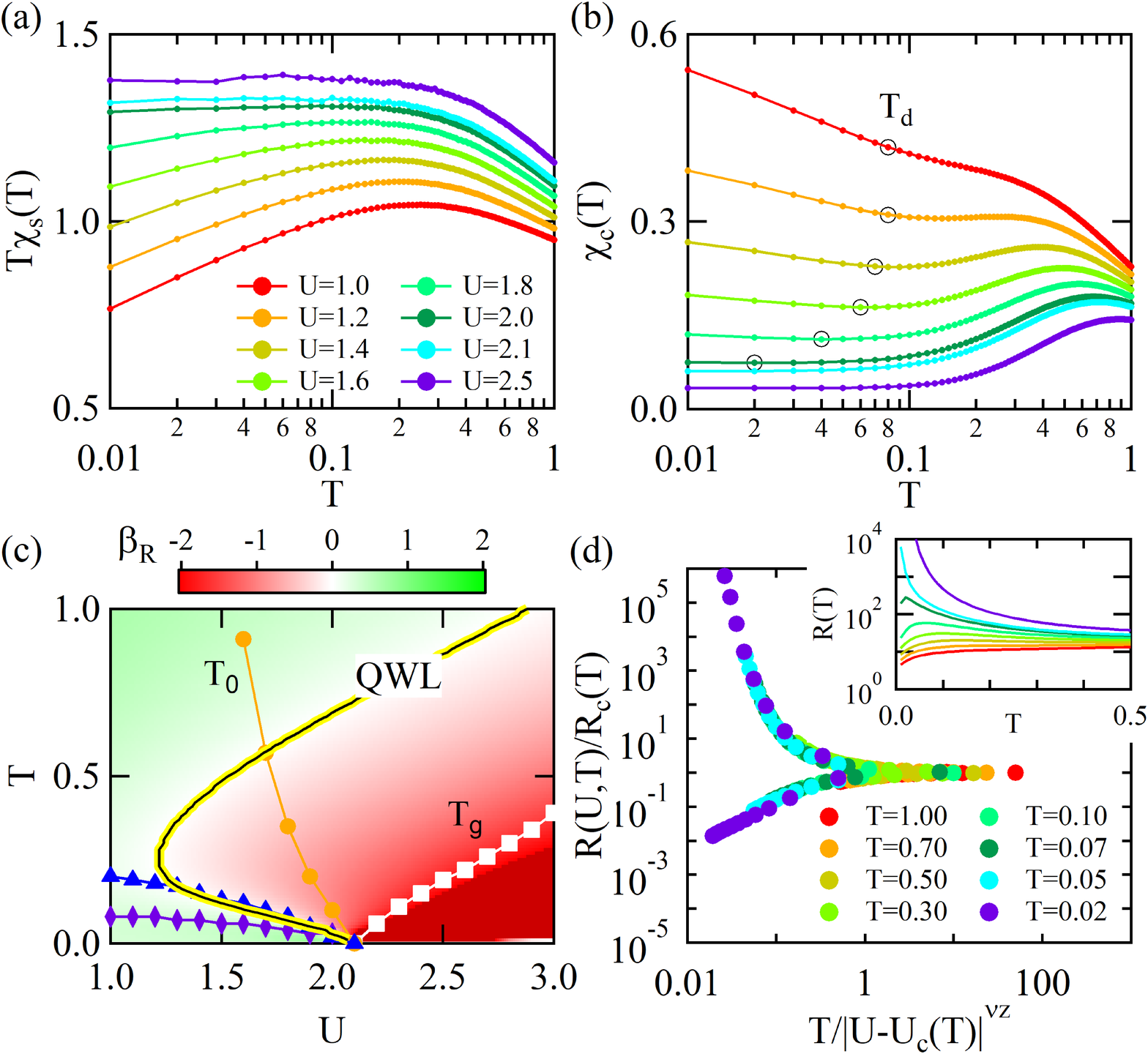}
	\caption{Temperature dependence of (a) the spin susceptibility $T\chi_s$ and (b) the charge susceptibility $\chi_c$ for different values of $U$. The black circles in (b) marks the temperature $T_d$ where the doublon (holon) occupation is lowest for each $U<U_c$. (c) Intensity plot of the temperature derivative of the resistivity, $\beta_R(U,T)\equiv\text{d}\log R(U,T)/\text{d}\log T$ on the $T$-$U$ plane, showing all four temperature scales $T_d$, $T_q$, $T_0$, and $T_g$. A quantum Widom line (QWL) can be identified from $\beta_R(U,T)=0$. (d) Quantum critical scaling of the resistivity, $R(U,T)/R_c(T)=f(T/|U-U_c(T)|^{\nu z})$, where $R_c(T)\equiv R(U_c(T),T)$, with the exponent $\nu z\approx1.0$. The inset of (d) shows the original data of the resistivity.}
	\label{fig3}
\end{figure}

Next we consider thermodynamic and transport properties within our method. Figures \ref{fig3}(a) and \ref{fig3}(b) plot the spin and charge susceptibility calculated using a bubble diagram of the slave particles:
\begin{eqnarray}\label{susceptibility}
\chi_s&=&\;4\int_{-\infty}^\infty\frac{\text d\omega}{\pi}n_B(\omega)\text{Re}G_s(\omega)\text{Im}G_s(\omega),\notag\\ 
\chi_c&=&\;4\int_{-\infty}^\infty\frac{\text d\omega}{\pi}n_F(\omega)\text{Re}G_d(\omega)\text{Im}G_d(\omega),
\end{eqnarray}
where $n_F(\omega)$ is the Fermi-Dirac distribution. For large $U$, $T\chi_s(T)$ approaches a constant at low temperatures, indicating that the spinons form well-defined local moments in the insulating phase. Its decrease at small $U$ suggests that the moments are partially screened in the metallic phase, consistent with the expectation that the spinons are damped for small $U$. The behavior of the charge susceptibility $\chi_c(T)$ in the metallic phase is similar to that of DMFT \cite{Mazitov2022DMFTChargeSusceptibility}. As $T$ decreases, $\chi_c$ first increases due to the development of quasiparticle coherence, then decreases due to the formation of local moments, and finally increases again due to the screening of local moments. As expected, the onset temperature of the final increase in $\chi_c$ is close to $T_d$ (black circles), where $n_d(T)$ has a local minimum but starts to grow at lower temperatures.

The resistivity $R=1/\sigma$ is calculated using the following equation \cite{DMFTReview1996}:
\begin{eqnarray}\label{resistance}
\sigma=-\frac{4t^2}{\mathcal N}\sum_{\bm k,\mu}\sin^2(k_\mu)\int^\infty_{-\infty}\frac{\text d\omega}{\pi}\frac{\partial n_F(\omega)}{\partial\omega}[\text{Im}G_c(\bm k,\omega)]^2.
\end{eqnarray}
The data are presented in the inset of Fig. \ref{fig3}(d) for the same $U$ values as in Fig. \ref{fig3}(a). For $U<U_c$, $R$ displays a small maximum and then drops as $T$ goes down, while for $U>U_c$, it diverges at low temperatures in accordance with experimental observations in both organic compounds and TMDs \cite{Tan2022MottCriticalityReview}. Figure \ref{fig3}(c) gives the intensity plot of $\beta_R(U,T)\equiv\text{d}\log R(U,T)/\text{d}\log T$ on the $T$-$U$ phase diagram. Besides the three temperature scales $T_q$, $T_0$, $T_g$ already shown in Fig. \ref{fig2}(c), we also include $T_d$ in the metallic phase. Interestingly, we can identify a clear quantum Widom line (QWL) \cite{Vucicevic2013MottCriticality} $U_c(T)$ from $\beta_R(U,T)=0$, which has the same shape as that reported in organic compounds \cite{Pustogow2018MottPhaseDiagram}, and partly coincides with $T_q$ identified from the lower boundary of the pseudogap region. Around the QWL, as shown in Fig. \ref{fig3}(d), our calculated resistivity data all collapse onto two separate curves for the insulating and metallic phases, respectively. We find a critical scaling $R(U,T)/R[U_c(T),T]=f[T/|U-U_c(T)|^{\nu z}]$ around the QWL. A similar scaling has been reported previously in both experiment \cite{Furukawa2015MottCriticality} and DMFT calculations \cite{Terletska2011MottCriticality}. However, the critical exponent is found to be $\nu z\approx 1.0$ here, in contrast to $\nu z=0.67$ from DMFT and $0.5\sim0.7$ from organic compounds and TMDs \cite{Tan2022MottCriticalityReview}. Compared to DMFT, we suspect that vertex correction, which is ignored in the one-loop approximation, might be responsible for this difference.

\begin{figure}
	\includegraphics[width=8.6cm]{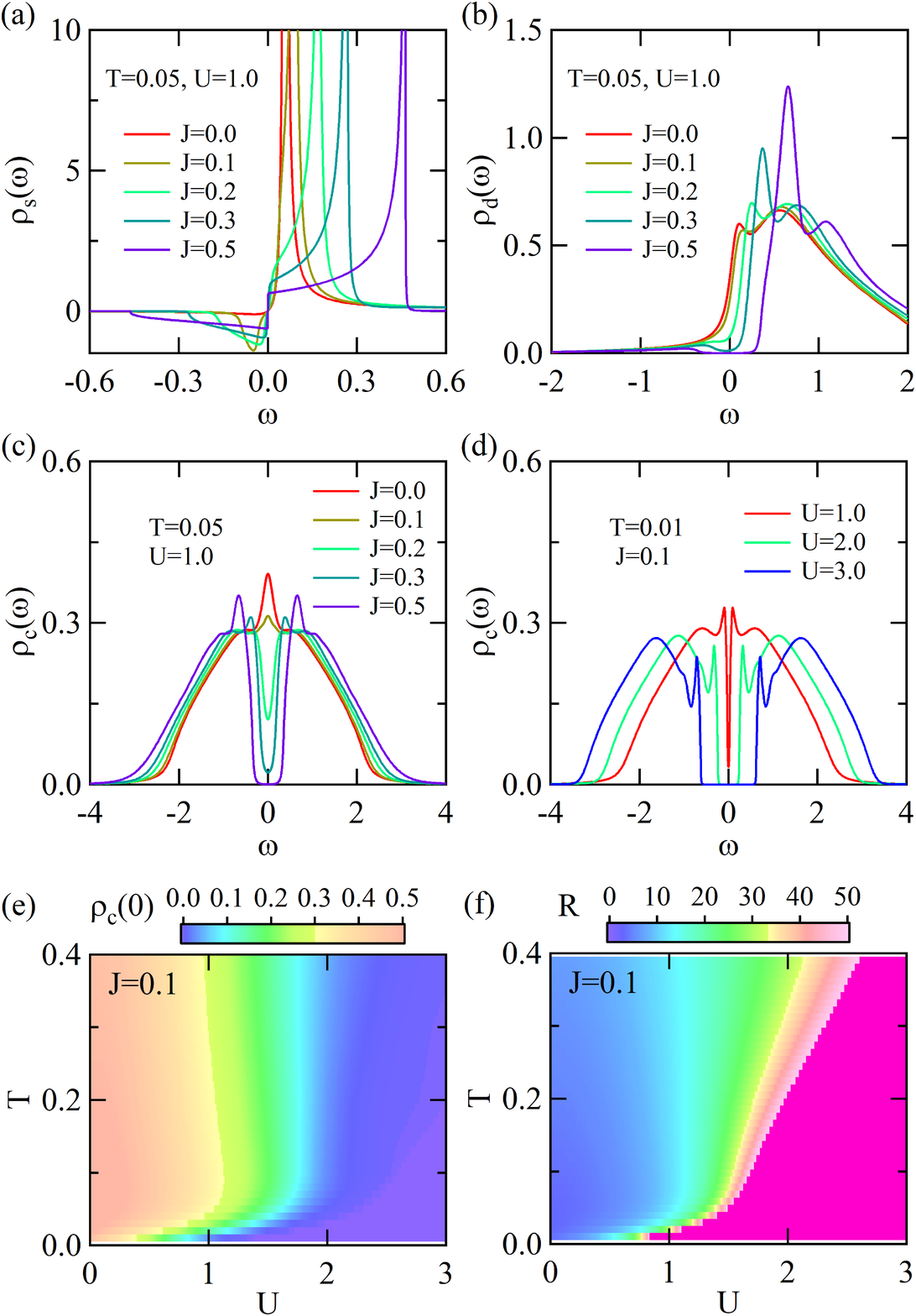}
	\caption{DOS of (a) spinon, (b) doublon (holon), and (c) electrons for different values of $J$ at $U=1.00$ and $T=0.05$. (d) Electron DOS for different values of $U$ at $J=0.1$ and $T=0.01$. (e),(f) Intensity plots of the electron DOS $\rho_c(0)$ at the Fermi energy and the resistivity $R$ on the $T$-$U$ plane.}
	\label{fig4}
\end{figure}

So far we have adopted a local approximation and ignored spatial correlation. To see its potential effects on the Mott transition \cite{Schafer2015AFM_and_Mott}, we include in the Hamiltonian an additional spinon-correlation term, $H_{\text{spin}}=J\sum_{\bm k}\xi_{\bm k}(s^\dagger_{\bm k,\uparrow}s^\dagger_{-\bm k,\downarrow}+\text{H.c.})$, which can be derived from the Heisenberg interaction, $J_H\sum_{\langle ij\rangle}\bm S_i\cdot\bm S_j\rightarrow -(J_H/2)\sum_{\langle ij\rangle}(A_{ij}s^\dagger_{i\uparrow}s^\dagger_{j\downarrow}+\text{H.c.})$, with $A_{ij}\equiv \sum_\sigma\langle\sigma s_{i\sigma}s_{j,-\sigma}\rangle$ under Schwinger boson mean-field approximation \cite{Arovas1988SchwingerBoson,Read1991SchwingerBoson}. For simplicity, we will not solve $A_{ij}$ self-consistently, but give by hand a dispersion $\xi_{\bm k}\equiv 2t[\sin(k_x)+\sin(k_y)]$. Again, we can solve the one-loop self-consistent equations for the local self-energies. The resulting spinon spectra are plotted in Fig. \ref{fig4}(a) for different values of $J$ at a fixed $T$ and $U=1.0$. We see two major features. First, the spinon peak shifts to higher energy linearly with increasing $J$, namely $\omega_s\sim J$. Second, for finite $J$, a sharp change occurs near $\omega=0$ at sufficiently low temperatures, which is a precursor of spinon condensation and implies an AFM ground state at zero temperature  \cite{Sachdev1992SchwingerBoson,Wang2020SchwingerBoson}. Meanwhile, as shown in Fig. \ref{fig4}(b), the doublon (holon) peak near $\omega=0$ gradually moves to higher energy, and a charge gap opens at low temperature even for small $U$. Correspondingly, the physical electron spectra in Fig. \ref{fig4}(c) also exhibit a gap near the Fermi energy. Figure \ref{fig4}(d) shows the electron spectra $\rho_c(\omega)$ for different $U$ at $J=0.1$. For all $U$, $\rho_c(\omega)$ always shows a gap and additional narrow peak structures on the inner side of the Hubbard bands at sufficiently low temperatures. These inner peaks were also captured in former QMC and slave fermion SCBA calculations \cite{Han2019SlaveFermion, Preuss1995FourPeakQMC}. To determine the transition point, we fix $J=0.1$ and give the intensity plots of $\rho_c(0)$ in Fig. \ref{fig4}(e) and the resistivity $R$ in Fig. \ref{fig4}(f) on the $T$-$U$ plane. We see quite similar behaviors on both plots, where $\rho_c(0)$ always decreases and $R$ diverges as the temperature approaches zero. These confirm an insulating ground state for all $U>0$, although we cannot perform calculations at sufficiently low temperature to confirm the gap opening for very small $U$. Thus, intersite AFM correlations drive the metal-insulator transition to $U=0$, as predicted in exact DQMC calculations. 
However, a better estimate of $J$ will be needed in order to obtain good prediction of the gap size compared to the exact QMC results \cite{Vitali2016U=0gap}.

The above spectral features have also been discussed previously in the literature and ascribed to doublon-holon binding or doublon-spinon interaction \cite{Castellani1979DHBinding, Kaplan1982DHBinding, Capello2005DHBinding, Yokoyama2006DHBinding, Zhou2014DHBinding, Sato2014DHBinding, Prelovsek2015DHBinding, Han2016SlaveFermion, Zhou2020DHBinding, MottReview2010,Terashige2019DHbindingExperiment, Han2019SlaveFermion,Schmitt-Rink1988SpinPolaron,Kane1988SpinPolaron,Martinez1988SpinPolaron}. In our method, their origins are quite transparent since these features only appear after we introduce the spinon correlation term, so they must be associated with the interaction of doublons (holons) with a correlated spinon background. In fact, if we again integrate out the spinon fields, the new spinon-correlation term will induce an effective pairing of the form $d_i e_j$, which causes the doublon-holon binding, produces an additional contribution to the charge gap in the doublon (holon) excitation spectra even for small $U$, and drives the metal-insulator transition to $U=0$. On the other hand, for large $U$ or $J$, we find the inner peak appears only at very low temperatures when the spinon gap approaches zero, which confirms its ``polaronic" origin \cite{Schmitt-Rink1988SpinPolaron,Kane1988SpinPolaron,Martinez1988SpinPolaron} as the doublons (holons) move on a magnetically correlated background \cite{Han2016SlaveFermion,Han2019SlaveFermion}. Interestingly, we also find that the Mott gap in the electron spectra is always smaller than the charge gap in the doublon/holon spectra at finite temperatures and they only become equal at zero temperature. This feature has also been observed in previous SCBA calculations \cite{Han2019SlaveFermion} and may be attributed to the convolution of the doublon/holon and spinon spectra in calculating the electron spectra. It reflects an intrinsic spin-charge separation nature of the Mott physics, while the electron quasiparticle can only be excited as a composite object of holon/doublon and spinon. For small $U$ and $J$, the hybridization effect still plays a role and is of primary responsibility for the peak, which is different from the situation at large $U$. 

\section{Conclusion}
To summarize, we develop a novel approach to study the mechanism of Mott metal-insulator transition. Our method combines the slave fermion representation of the electrons and the fermionic auxiliary field to decouple the kinetic term in the Hamiltonian. This results in a three-particle vertex that may be solved by a self-consistent one-loop approximation. Our calculations reproduce the Mott transition as a continuous quantum phase transition at zero temperature and associate the quasiparticle emergence with a dynamic charge Kondo effect. We also derive a phase diagram containing a pseudogap region above the metallic phase and obtain a resistivity scaling around the quantum Widom line. Including AFM spin correlation  drives the Mott transition to $U=0$ and induces some special spectral features, confirming the importance of intersite magnetic correlations. Hence, our theory captures the essential features of the Mott physics and provides an alternative angle for clarifying this long-standing problem. Our method may be easily extended to cover a large family of correlated models with more complex local interactions.

It may be helpful to compare our method with other slave particle approaches such as the Kotliar-Ruckenstein slave boson \cite{Kotliar1986KRSlaveBoson} and the slave rotor \cite{Florens2004SlaveRotor} approaches. There are three major differences. First, in the slave boson approaches, doublons and holons are treated as bosons and the metallic phase is realized when they condense. Under mean-field approximation, it is equivalent to the Gutzwiller approach and can only account for the quasiparticle bands near the Fermi energy. To obtain the Hubbard bands, additional Gaussian fluctuations \cite{Raimondi1993GaussianofKR} or a more complicated approximation \cite{Zhou2014DHBinding} are needed. Similarly, in the slave rotor approach, the charge degree of freedom is described as a rotor and the metallic phase is obtained by the condensation of a constrained bosonic field  \cite{Florens2004SlaveRotor}. In our slave fermion approach, the Hubbard bands come naturally from the fermionic doublon/holon states, and the Mott transition is associated with a dynamic charge Kondo effect tuned by the doublon/holon levels. While it seems less intuitive, the Kondo effect does not involve any symmetry breaking and the transition takes place likely between two orthogonal ground states at zero temperature \cite{HewsonBook}. Second, in slave boson and slave rotor approaches, the spins are described by fermions. Because the mean-field approximation replaces the local constraint by lattice average, the fermionic representation of spins is more appropriate for describing a fermionic spin liquid with a spinon Fermi surface rather than long-range magnetic orders of localized moments. By contrast, the bosonic representation of the spin degree of freedom can describe well the magnetic long-range order of local moments by the condensation of bosonic spinons. Third, in all above approaches, the kinetic term in the Hamiltonian becomes quite complicated under the slave particle representation, so a mean-field approximation has often been applied to decouple the kinetic term. By contrast, our work introduces an additional fermionic auxiliary field to describe the kinetic term, and thereby successfully applies the one-loop self-consistent calculations of the Green's functions. This naturally takes into account some dynamic fluctuations of the slave particles beyond the mean-field approximation. It will be interesting to see if our idea can also be applied to the bosonic representations and yield improved description of the Mott transition there.

\textit{Note added}: After completing this work, we were informed of a simultaneous study by DMFT \cite{Mazitov2022DMFTResistivity}, which reported some similar results as in Fig. 3 including the coincidence of the QWL and $T_q$ at low temperatures. The agreement again confirms the validity of our slave fermion approach.

\acknowledgements
This work was supported by the National Natural Science Foundation of China (Grants No. 12174429, No. 11974397), and the Strategic Priority Research Program of the Chinese Academy of Sciences (Grant No. XDB33010100).


\begin{thebibliography}{150}
\bibitem{Hubbard1963}  J. Hubbard, 
Electron Correlations in Narrow Energy Bands, 
Proc. R. Soc. London A {\bf 276}, 238 (1963).

\bibitem{Brinkman-Rice1970}  W. Brinkman and T. M. Rice, 
Application of Gutzwiller's Variational Method to the Metal-Insulator Transition, 
Phys. Rev. B {\bf 2}, 4302 (1970).

\bibitem{MetalInsulatorTransitionsReview1998} M. Imada, A. Fujimori, and Y. Tokura, 
Metal-insulator transitions, 
Rev. Mod. Phys. {\bf 70}, 1039 (1998).

\bibitem{Dobrosavljevic2012Book} V. Dobrosavljevi\'{c}, N. Trivedi, and J. M. Valles, Jr., 
\textit{Conductor-Insulator Quantum Phase Transitions} 
(Oxford University Press, Oxford, England, 2012).

\bibitem{Phillips2022Z2symmetry} E. W. Huang, G. L. Nave, and P. W. Phillips, 
Discrete symmetry breaking defines the Mott quartic fixed point, 
Nat. Phys. {\bf 18}, 511 (2022).

\bibitem{McWhan1973V2O3PhaseDigram} D. B. McWhan, A. Menth, J. P. Remeika, W. F. Brinkman, and T. M. Rice, 
Metal-Insulator Transitions in Pure and Doped ${\mathrm{V}}_{2}{\mathrm{O}}_{3}$, 
Phys. Rev. B {\bf 7}, 1920 (1973).

\bibitem{Limelette2003V2O3} P. Limelette, A. Georges, D. J\'{e}rome, P. Wzietek, P. Metcalf, and J. M. Honig, 
Universality and Critical Behavior at the Mott Transition, 
Science {\bf 302}, 89 (2003).

\bibitem{Lefebvre2000kappaCl} S. Lefebvre, P. Wzietek, S. Brown, C. Bourbonnais, D. J\'{e}rome, C. M\'{e}zi\`{e}re, M. Fourmigu\'{e}, and P. Batail,  
Mott Transition, Antiferromagnetism, and Unconventional Superconductivity in Layered Organic Superconductors, 
Phys. Rev. Lett. {\bf 85}, 5420 (2000).

\bibitem{Kagawa2005kappaCl_Scaling} F. Kagawa, K. Miyagawa, and K. Kanoda, 
Unconventional critical behaviour in a quasi-two-dimensional organic conductor, 
Nature {\bf 436}, 534 (2005).

\bibitem{Kurosaki2005kappaCu2(CN)3} Y. Kurosaki, Y. Shimizu, K. Miyagawa, K. Kanoda, and G. Saito, 
Mott Transition from a Spin Liquid to a Fermi Liquid in the Spin-Frustrated Organic Conductor $\ensuremath{\kappa}\text{-}(\mathrm{ET})_{2}{\mathrm{Cu}}_{2}(\mathrm{CN})_{3}$, 
Phys. Rev. Lett. {\bf 95}, 177001 (2005).

\bibitem{Furukawa2018kappaCu2(CN)3} T. Furukawa, K. Kobashi, Y. Kurosaki, K. Miyagawa, and K. Kanoda, 
Quasi-continuous transition from a Fermi liquid to a spin liquid in  $\ensuremath{\kappa}\text{-}(\mathrm{ET})_{2}{\mathrm{Cu}}_{2}(\mathrm{CN})_{3}$, 
Nat. Commun. {\bf 9}, 307 (2018).

\bibitem{Ghiotto2021TMD} A. Ghiotto, E.-M. Shih, G. S. S. G. Pereira, D. A. Rhodes, B. Kim, J. Zang, A. J. Millis, K. Watanabe, T. Taniguchi, J. C. Hone, L. Wang, C. R. Dean, and A. N. Pasupathy,  
Quantum criticality in twisted transition metal dichalcogenides,
Nature {\bf 597}, 345 (2021).

\bibitem{Li2021TMD} T. Li, S. Jiang, L. Li, Y. Zhang, K. Kang, J. Zhu, K. Watanabe, T. Taniguchi, D. Chowdhury, L. Fu, J. Shan, and K. F. Mak, 
Continuous Mott transition in semiconductor moiré superlattices,
Nature {\bf 597}, 350 (2021).

\bibitem{Mo2003V2O3PES} S.-K. Mo, J. D. Denlinger, H.-D. Kim, J.-H. Park, J. W. Allen, A. Sekiyama, A. Yamasaki, K. Kadono, S. Suga, Y. Saitoh, T. Muro, P. Metcalf, G. Keller, K. Held, V. Eyert, V. I. Anisimov, and D. Vollhardt, 
Prominent Quasiparticle Peak in the Photoemission Spectrum of the Metallic Phase of ${\mathrm{V}}_{2}{\mathrm{O}}_{3}$, 
Phys. Rev. Lett. {\bf 90}, 186403 (2003).

\bibitem{Xu2014NiS2-xSexARPES} H. C. Xu, Y. Zhang, M. Xu, R. Peng, X. P. Shen, V. N. Strocov, M. Shi, M. Kobayashi, T. Schmitt, B. P. Xie, and D. L. Feng, 
Direct Observation of the Bandwidth Control Mott Transition in the ${\mathrm{NiS}}_{2\text{-}x}{\mathrm{Se}}_{x}$ Multiband System, 
Phys. Rev. Lett. {\bf 112}, 087603 (2014).

\bibitem{Furukawa2015MottCriticality} T. Furukawa, K. Miyagawa, H. Taniguchi, R. Kato, and K. Kanoda, 
Quantum criticality of Mott transition in organic materials,
Nat. Phys. {\bf 11}, 221 (2015).

\bibitem{Arovas2022HubbardReview}  D. P. Arovas, E. Berg, S. A. Kivelson, and S. Raghu,
The Hubbard Model,
Annu. Rev. Condens. Matter Phys. {\bf 13}, 239 (2022).

\bibitem{Qin2022NumericalReview}  M. Qin, T. Sch\"{a}fer, S. Andergassen, P. Corboz, and E. Gull,
The Hubbard Model: A Computational Perspective,
Annu. Rev. Condens. Matter Phys. {\bf 13}, 275 (2022).

\bibitem{Kotliar1986KRSlaveBoson} G. Kotliar and A. E. Ruckenstein, 
New Functional Integral Approach to Strongly Correlated Fermi Systems: The Gutzwiller Approximation as a Saddle Point, 
Phys. Rev. Lett. {\bf 57}, 1362 (1986).

\bibitem{Li1989SRIKR} T. Li, P. W\"{o}lfle, and P. J. Hirschfeld, 
Spin-rotation-invariant slave-boson approach to the Hubbard model, 
Phys. Rev. B {\bf 40}, 6817 (1989).

\bibitem{Fresard1992SRIKR} R. Fr\'{e}sard and P. W\"{o}lfle, 
Unified Slave Boson Representation of Spin and Charge Degrees of Freedom for Strongly Correlated Fermi Systems,
Int. J. Mod. Phys. B {\bf 6}, 685 (1992).

\bibitem{CupratesReview2006} P. A. Lee, N. Nagaosa, and X.-G. Wen, 
Doping a Mott insulator: Physics of high-temperature superconductivity,
Rev. Mod. Phys. {\bf 78}, 17 (2006).  

\bibitem{Kaga1992SlaveBosonCompare} H. Kaga, 
Finite-$U$ slave-boson methods and the intersite magnetic correlations in the Hubbard models, 
Phys. Rev. B {\bf 46}, 1979 (1992).

\bibitem{Florens2002SlaveRotor} S. Florens and A. Georges, 
Quantum impurity solvers using a slave rotor representation,
Phys. Rev. B {\bf 66}, 165111 (2002).

\bibitem{Florens2004SlaveRotor} S. Florens and A. Georges, 
Slave-rotor mean-field theories of strongly correlated systems and the Mott transition in finite dimensions,
Phys. Rev. B {\bf 70}, 035114 (2004).

\bibitem{Lee2005SpinLiquid} S.-S. Lee and P. A. Lee, 
U(1) Gauge Theory of the Hubbard Model: Spin Liquid States and Possible Application to $\ensuremath{\kappa}\text{-}(\mathrm{BEDT}\text{-}\mathrm{TTF})_{2}{\mathrm{Cu}}_{2}(\mathrm{CN})_{3}$,
Phys. Rev. Lett. {\bf 95}, 036403 (2005).

\bibitem{Kim2006SU(2)SlaveRotor} K.-S. Kim, 
SU(2) Gauge Theory of the Hubbard Model: Emergence of an Anomalous Metallic Phase near the Mott Critical Point,
Phys. Rev. Lett. {\bf 97}, 136402 (2006).

\bibitem{Hermele2007SU(2)SlaveRotor} M. Hermele, 
SU(2) gauge theory of the Hubbard model and application to the honeycomb lattice,
Phys. Rev. B {\bf 76}, 035125 (2007).

\bibitem{Zhao2007SlaveRotorClusterMFT} E. Zhao and A. Paramekanti, 
Self-consistent slave rotor mean-field theory for strongly correlated systems,
Phys. Rev. B {\bf 76}, 195101 (2007).

\bibitem{Senthil2008CriticalFS} T. Senthil, 
Critical Fermi surfaces and non-Fermi liquid metals,
Phys. Rev. B {\bf 78}, 035103 (2008).

\bibitem{Senthil2008SlaveRotor} T. Senthil, 
Theory of a continuous Mott transition in two dimensions,
Phys. Rev. B {\bf 78}, 045109 (2008).

\bibitem{Podolsky2009SlaveRotor3DCase} D. Podolsky, A. Paramekanti, Y. B. Kim, and T. Senthil, 
Mott Transition between a Spin-Liquid Insulator and a Metal in Three Dimensions,
Phys. Rev. Lett. {\bf 102}, 186401 (2009).

\bibitem{Xu2022TMDMottTransition} Y. Xu, X.-C. Wu, M. Ye, Z.-X. Luo, C.-M. Jian, and C. Xu,
Interaction-Driven Metal-Insulator Transition with Charge Fractionalization,
Phys. Rev. X {\bf 12}, 021067 (2022)

\bibitem{de'Medici2005Z2SlaveSpin1} L. de' Medici, A. Georges, and S. Biermann, 
Orbital-selective Mott transition in multiband systems: Slave-spin representation and dynamical mean-field theory,
Phys. Rev. B {\bf 72}, 205124 (2005).

\bibitem{de'Medici2010Z2SlaveSpin1} S. R. Hassan and L. de' Medici, 
Slave spins away from half filling: Cluster mean-field theory of the Hubbard and extended Hubbard models,
Phys. Rev. B {\bf 81}, 035106 (2010).

\bibitem{Ruegg2010Z2SlaveSpin2} A. R\"{u}egg, S. D. Huber, and M. Sigrist, 
${\mathsf{Z}}_{2}$-slave-spin theory for strongly correlated fermions,
Phys. Rev. B {\bf 81}, 155118 (2010).

\bibitem{Nandkishore2010OrthogonalMetal} R. Nandkishore, M. A. Metlitski, and T. Senthil, 
Orthogonal metals: The simplest non-Fermi liquids,
Phys. Rev. B {\bf 86}, 045128 (2012).

\bibitem{Zitko2015MottZ2} R. \v{Z}itko and M. Fabrizio,
${Z}_{2}$ gauge theory description of the Mott transition in infinite dimensions,
Phys. Rev. B {\bf 91}, 245130 (2015).

\bibitem{Yu2012U(1)SlaveSpin} R. Yu and Q. Si,
$U(1)$ slave-spin theory and its application to Mott transition in a multiorbital model for iron pnictides,
Phys. Rev. B {\bf 86}, 085104 (2012).

\bibitem{Yu2017U(1)SlaveSpin} R. Yu and Q. Si,
Orbital-selective Mott phase in multiorbital models for iron pnictides and chalcogenides,
Phys. Rev. B {\bf 96}, 125110 (2017).

\bibitem{Castellani1979DHBinding}  C. Castellani, C. D. Castro, D. Feinberg, and J. Ranninger, 
New Model Hamiltonian for the Metal-Insulator Transition,
Phys. Rev. Lett. {\bf 43}, 1957 (1979).

\bibitem{Kaplan1982DHBinding}  T. A. Kaplan, P. Horsch, and P. Fulde, 
Close Relation between Localized-Electron Magnetism and the Paramagnetic Wave Function of Completely Itinerant Electrons,
Phys. Rev. Lett. {\bf 49}, 889 (1982).

\bibitem{Capello2005DHBinding}  M. Capello, F. Becca, M. Fabrizio, S. Sorella, and E. Tosatti,
Variational Description of Mott Insulators,
Phys. Rev. Lett. {\bf 94}, 026406 (2005).

\bibitem{Yokoyama2006DHBinding}  H. Yokoyama, M. Ogata, and Y. Tanaka,
Mott Transitions and d-Wave Superconductivity in Half-Filled-Band Hubbard Model on Square Lattice with Geometric Frustration,
J. Phys. Soc. Jpn. {\bf 75}, 114706 (2006).

\bibitem{MottReview2010}  P. W. Phillips, 
Colloquium: Identifying the propagating charge modes in doped Mott insulators,
Rev. Mod. Phys. {\bf 82}, 1719 (2010).

\bibitem{Zhou2014DHBinding}  S. Zhou, Y. Wang, and Z. Wang, 
Doublon-holon binding, Mott transition, and fractionalized antiferromagnet in the Hubbard model,
Phys. Rev. B {\bf 89}, 195119 (2014).

\bibitem{Sato2014DHBinding}  T. Sato and H. Tsunetsugu,
Doublon dynamics of the Hubbard model on a triangular lattice,
Phys. Rev. B {\bf 90}, 115114 (2014).

\bibitem{Prelovsek2015DHBinding} P. Prelov\v{s}ek, J. Kokalj, Z. Lenar\v{c}i\v{c}, and R. H. McKenzie, 
Holon-doublon binding as the mechanism for the Mott transition,
Phys. Rev. B {\bf 92}, 235155 (2015).

\bibitem{Han2016SlaveFermion} X.-J. Han, Y. Liu, Z.-Y. Liu, X. Li, J. Chen, H.-J. Liao, Z.-Y. Xie, B. Normand, and T. Xiang, 
Charge dynamics of the antiferromagnetically ordered Mott insulator,
New J. Phys. {\bf 18}, 103004 (2016).

\bibitem{Han2019SlaveFermion} X.-J. Han, C. Chen, J. Chen, H.-D. Xie, R.-Z. Huang, H.-J. Liao, B. Normand, Z. Y. Meng, and T. Xiang, 
Finite-temperature charge dynamics and the melting of the Mott insulator,
Phys. Rev. B {\bf 99}, 245150 (2019).

\bibitem{Zhou2020DHBinding}  S. Zhou, L. Liang, and Z. Wang, 
Dynamical slave-boson mean-field study of the Mott transition in the Hubbard model in the large-$z$ limit,
Phys. Rev. B {\bf 101}, 035106 (2020).

\bibitem{Leigh2007Charge2eBoson} R. G. Leigh, P. W. Phillips, and T.-P. Choy, 
Hidden Charge $2e$ Boson in Doped Mott Insulators,
Phys. Rev. Lett. {\bf 99}, 046404 (2007).

\bibitem{Leigh2009Charge2eBoson} R. G. Leigh and P. W. Phillips,
Origin of the Mott gap,
Phys. Rev. B {\bf 79}, 245120 (2009).

\bibitem{HKmodel1992} Y. Hatsugai and M. Kohmoto,  
Exactly solvable model of correlated lattice electrons in any dimensions, 
J. Phys. Soc. Jpn. {\bf 61}, 2056 (1992).

\bibitem{Phillips2020superconductivity}  P. W. Phillips, L. Yeo, and E. W. Huang, 
Exact theory for superconductivity in a doped Mott insulator,
Nat. Phys. {\bf 16}, 1175 (2020).

\bibitem{DMFTReview1996}  A. Georges, G. Kotliar, W. Krauth, and M. J. Rozenberg, 
Dynamical mean-field theory of strongly correlated fermion systems and the limit of infinite dimensions,
Rev. Mod. Phys. {\bf 68}, 13 (1996).

\bibitem{clusterDMFTReview2005}  T. Maier, M. Jarrell, T. Pruschke, and M. H. Hettler,
Quantum cluster theories,
Rev. Mod. Phys. {\bf 77}, 1027 (2005).

\bibitem{CTQMCReview2011}  E. Gull, A. J. Millis, A. I. Lichtenstein, A. N. Rubtsov, M. Troyer, and P. Werner,
Continuous-time Monte Carlo methods for quantum impurity models,
Rev. Mod. Phys. {\bf 83}, 349 (2011).

\bibitem{diagrammaticReview2018} G. Rohringer, H. Hafermann, A. Toschi, A. A. Katanin, A. E. Antipov, M. I. Katsnelson, A. I. Lichtenstein, A. N. Rubtsov, and K. Held,
Diagrammatic routes to nonlocal correlations beyond dynamical mean field theory,
Rev. Mod. Phys. {\bf 90}, 025003 (2018).

\bibitem{LeBlanc2015Benchmark} J. P. F. LeBlanc, A. E. Antipov, F. Becca, I. W. Bulik, G. K.-L. Chan, C.-M. Chung, Y. Deng, M. Ferrero, T. M. Henderson, C. A. Jim\'{e}nez-Hoyos. E. Kozik. X.-W. Liu, A. J. Millis, N. V. Prokof'ev, M. Qin, G. E. Scuseria, H. Shi, B. V. Svistunov, L. F. Tocchio, I. S. Tupitsyn, S. R. White, S. Zhang, B.-X. Zheng, Z. Zhu, and E. Gull,
Solutions of the Two-Dimensional Hubbard Model: Benchmarks and Results from a Wide Range of Numerical Algorithms,
Phys. Rev. X {\bf 5}, 041041 (2015).

\bibitem{Schafer2021Benchmark} T. Sch\"{a}fer, N. Wentzell, F. \v{S}imkovic, Y.-Y. He, C. Hille, M. Klett, C. J. Eckhardt, B. Arzhang, V. Harkov, F.-M. Le R\'{e}gent, A. Kirsch, Y. Wang, A. J. Kim, E. Kozik, E. A. Stepanov, A. Kauch, S. Andergassen, P. Hansmann, D. Rohe, Y. M. Vilk, J. P. F. LeBlanc, S. Zhang, A.-M. S. Tremblay, M. Ferrero, O. Parcollet, and A. Georges,
Tracking the Footprints of Spin Fluctuations: A MultiMethod, MultiMessenger Study of the Two-Dimensional Hubbard Model,
Phys. Rev. X {\bf 11}, 011058 (2021).

\bibitem{Meinders1993DSWT} M. B. J. Meinders, H. Eskes, and G. A. Sawatzky,
Spectral-weight transfer: Breakdown of low-energy-scale sum rules in correlated systems,
Phys. Rev. B {\bf 48}, 3916 (1993).

\bibitem{Eskes1994DSWT} H. Eskes, A. M. Ole\'{s}, M. B. J. Meinders, and W. Stephan,
Spectral properties of the Hubbard bands,
Phys. Rev. B {\bf 50}, 17980 (1994).

\bibitem{Dzyaloshinskii2003LuttingerTheorem}   I. Dzyaloshinskii, 
Some consequences of the Luttinger theorem: The Luttinger surfaces in non-Fermi liquids and Mott insulators, 
Phys. Rev. B {\bf 68}, 085113 (2003).

\bibitem{Borejsza2004NLSM} K. Borejsza and N. Dupuis, 
Antiferromagnetism and single-particle properties in the two-dimensional half-filled Hubbard model: A nonlinear sigma model approach,
Phys. Rev. B {\bf 69}, 085119 (2004).

\bibitem{Held2000PAMandHubbard}  K. Held, C. Huscroft, R. T. Scalettar, and A. K. McMahan,
Similarities between the Hubbard and Periodic Anderson Models at Finite Temperatures,
Phys. Rev. Lett. {\bf 85}, 373 (2000).

\bibitem{Hu2020ClassicalCorrespondence}   D. Hu, J.-J. Dong, L. Huang, L. Wang, and Y.-F. Yang, 
Effective classical correspondence of the Mott transition, 
Phys. Rev. B {\bf 101}, 075111 (2020).

\bibitem{Hu2020PAMandHubbard}   D. Hu, N.-H. Tong, and Y.-F. Yang, 
Energy-scale cascade and correspondence between Mott and Kondo lattice physics, 
Phys. Rev. Research {\bf 2}, 043407 (2020).

\bibitem{Byczuk2012RelativeEntropy}   K. Byczuk, J. Kune\v{s}, W. Hofstetter, and D. Vollhardt,
Quantification of Correlations in Quantum Many-Particle Systems,
Phys. Rev. Lett. {\bf 108}, 087004 (2012).

\bibitem{Walsh2019Information} C. Walsh, P. S\'{e}mon, D. Poulin, G. Sordi, and A.-M. S.  Tremblay,
Local Entanglement Entropy and Mutual Information across the Mott Transition in the Two-Dimensional Hubbard Model,
Phys. Rev. Lett. {\bf 122}, 067203 (2019).

\bibitem{Sen2020Topological}   S. Sen, P. J. Wong, and A. K. Mitchell,
The Mott transition as a topological phase transition,
Phys. Rev. B {\bf 102}, 081110(R) (2020).

\bibitem{Terletska2011MottCriticality}  H. Terletska, J. Vu\v{c}i\v{c}evi\'{c}, D. Tanaskovi\'{c}, and V. Dobrosavljevi\'{c}, 
Quantum Critical Transport near the Mott Transition,
Phys. Rev. Lett. {\bf 107}, 026401 (2011).

\bibitem{Vucicevic2013MottCriticality}  J. Vu\v{c}i\v{c}evi\'{c}, H. Terletska, D. Tanaskovi\'{c}, and V. Dobrosavljevi\'{c}, 
Finite-temperature crossover and the quantum Widom line near the Mott transition,
Phys. Rev. B {\bf 88}, 075143 (2013).

\bibitem{Vucicevic2015MottCriticality}  J. Vu\v{c}i\v{c}evi\'{c}, D. Tanaskovi\'{c}, M. J. Rozenberg, and V. Dobrosavljevi\'{c},
Bad-Metal Behavior Reveals Mott Quantum Criticality in Doped Hubbard Models,
Phys. Rev. Lett. {\bf 114}, 246402 (2015).

\bibitem{Lee2016MottCriticalityAndSpinon}  T.-H. Lee, S. Florens, and V. Dobrosavljevi\'{c},
Fate of Spinons at the Mott Point,
Phys. Rev. Lett. {\bf 117}, 136601 (2016).

\bibitem{Dasari2017MottSusceptibilityScaling} N. Dasari, N. S. Vidhyadhiraja, M. Jarrell, and R. H. McKenzie,
Quantum critical local spin dynamics near the Mott metal-insulator transition in infinite dimensions,
Phys. Rev. B {\bf 95}, 165105 (2017).

\bibitem{Eisenlohr2019MottCriticality} H. Eisenlohr, S.-S. B. Lee, and M. Vojta,
Mott quantum criticality in the one-band Hubbard model: Dynamical mean-field theory, power-law spectra, and scaling,
Phys. Rev. B {\bf 100}, 155152 (2019).

\bibitem{Kotliar1999DMFTLandauTheory}  G. Kotliar,
Landau theory of the Mott transition in the fully frustrated Hubbard model in infinite dimensions,
Eur. Phys. J. B {\bf 11}, 27 (1999).

\bibitem{Kotliar2000DMFTLandauTheory}  G. Kotliar, E. Lange, and M. J. Rozenberg,
Landau Theory of the Finite Temperature Mott Transition,
Phys. Rev. Lett. {\bf 84}, 5180 (2000).

\bibitem{Hirsch1985DQMC}  J. E. Hirsch,
Two-dimensional Hubbard model: Numerical simulation study,
Phys. Rev. B {\bf 31}, 4403 (1985).

\bibitem{White1989DQMC}  S. R. White, D. J. Scalapino, R. L. Sugar, E. Y. Loh, J. E. Gubernatis, and R. T. Scalettar,
Numerical study of the two-dimensional Hubbard model,
Phys. Rev. B {\bf 40}, 506 (1989).

\bibitem{Preuss1995FourPeakQMC}  R. Preuss, W. Hanke, and W. von der Linden,
Quasiparticle Dispersion of the 2D Hubbard Model: From an Insulator to a Metal,
Phys. Rev. Lett. {\bf 75}, 1344 (1995).

\bibitem{Schafer2015AFM_and_Mott} T. Sch\"{a}fer, F. Geles, D. Rost, G. Rohringer, E. Arrigoni, K. Held, N. Bl\"{u}mer, M. Aichhorn, and A. Toschi, 
Fate of the false Mott-Hubbard transition in two dimensions,
Phys. Rev. B {\bf 91}, 125109 (2015).

\bibitem{Vitali2016U=0gap} E. Vitali, H. Shi, M. Qin, and S. Zhang,
Computation of dynamical correlation functions for many-fermion systems with auxiliary-field quantum Monte Carlo,
Phys. Rev. B {\bf 94}, 085140 (2016).

\bibitem{Simkovic2020U=0gap} F. \v{S}imkovic, J. P. F. LeBlanc, A. J. Kim, Y. Deng, N. V.  Prokof'ev, B. V. Svistunov, and E. Kozik,
Extended Crossover from a Fermi Liquid to a Quasiantiferromagnet in the Half-Filled 2D Hubbard Model,
Phys. Rev. Lett. {\bf 124}, 017003 (2020).

\bibitem{Yoshioka1989SlaveFermion} D. Yoshioka, 
Slave-Fermion Mean Field Theory of the Hubbard Model,
J. Phys. Soc. Jpn. {\bf 58}, 1516 (1989).

\bibitem{Jayaprakash1989SlaveFermion} C. Jayaprakash, H. R. Krishnamurthy, and S. Sarker, Mean-field theory for the $t$-$J$ model, Phys. Rev. B {\bf 40}, 2610(R) (1989).

\bibitem{Kane1990SlaveFermion} C. L. Kane, P. A. Lee, T. K. Ng, B. Chakraborty, and N. Read, Mean-field theory of the spiral phases of a doped antiferromagnet, Phys. Rev. B {\bf 41}, 2653(R) (1990).

\bibitem{Chakraborty1990SlaveFermion} B. Chakraborty, N. Read, C. Kane, and P. A. Lee, Spiral phases and time-reversal-violating resonating-valence-bond states of doped antiferromagnets, Phys. Rev. B {\bf 42}, 4819(R) (1990).

\bibitem{Auerbach1991SlaveFermion} A. Auerbach and B. E. Larson, 
Doped antiferromagnet: The instability of homogeneous magnetic phases,
Phys. Rev. B {\bf 43}, 7800 (1991).

\bibitem{Li1992SlaveFermion} Y. M. Li, D. N. Sheng, Z. B. Su, and L. Yu, 
Schwinger-boson studies of the $t$-$J$ model: A self-consistent systematic expansion,
Phys. Rev. B {\bf 45}, 5428 (1992).

\bibitem{Igarashi1992SlaveFermion} J. Igarashi and P. Fulde,
Spiral phase of a doped antiferromagnet,
Phys. Rev. B {\bf 45}, 10419 (1992).

\bibitem{Khaliullin1993SlaveFermion}  G. Khaliullin and P. Horsch,
Doping dependence of long-range magnetic order in the $t$-$J$ model,
Phys. Rev. B {\bf 47}, 463 (1993).

\bibitem{Parcollet1997SchwingerBoson} O. Parcollet and A. Georges,
Transition from Overscreening to Underscreening in the Multichannel Kondo Model: Exact Solution at Large $\mathit{N}$,
Phys. Rev. Lett. {\bf 79}, 4665 (1997).

\bibitem{Coleman2005SchwingerBoson} P. Coleman, I. Paul, and J. Rech, 
Sum rules and Ward identities in the Kondo lattice,
Phys. Rev. B {\bf 72}, 094430 (2005).

\bibitem{Rech2006SchwingerBoson} J. Rech, P. Coleman, G. Zarand, and O. Parcollet, 
Schwinger Boson Approach to the Fully Screened Kondo Model,
Phys. Rev. Lett. {\bf 96}, 016601 (2006).

\bibitem{Komijani2018SchwingerBoson} Y. Komijani and P. Coleman, 
Model for a Ferromagnetic Quantum Critical Point in a 1D Kondo Lattice,
Phys. Rev. Lett. {\bf 120}, 157206 (2018).

\bibitem{Komijani2019SchwingerBoson} Y. Komijani and P. Coleman, 
Emergent Critical Charge Fluctuations at the Kondo Breakdown of Heavy Fermions,
Phys. Rev. Lett. {\bf 122}, 217001 (2019).

\bibitem{Wang2020SchwingerBoson} J. Wang, Y.-Y. Chang, C.-Y. Mou, S. Kirchner, and C.-H. Chung, 
Quantum phase transition in a two-dimensional Kondo-Heisenberg model: A dynamical Schwinger-boson large-$N$ approach,
Phys. Rev. B {\bf 102}, 115133 (2020).

\bibitem{Wang2021SchwingerBoson} J. Wang and Y.-F. Yang, 
Nonlocal Kondo effect and quantum critical phase in heavy-fermion metals,
Phys. Rev. B {\bf 104}, 165120 (2021).

\bibitem{Han2021SchwingerBoson} R. Han, D. Hu, J. Wang, and Y-F. Yang, 
Schwinger boson approach for the dynamical mean-field theory of the Kondo lattice,
Phys. Rev. B {\bf 104}, 245132 (2021).

\bibitem{Wang2022SchwingerBosonSpinCurrent} J. Wang and Y.-F. Yang, 
Spin current Kondo effect in frustrated Kondo systems,
Sci. China Phys. Mech. Astron. {\bf 65}, 227212 (2022).

\bibitem{Wang2022SchwingerBosonFMQCP} J. Wang and Y.-F. Yang, 
A unified theory of ferromagnetic quantum phase transitions in heavy fermion metals,
Sci. China Phys. Mech. Astron. {\bf 65}, 257211 (2022).

\bibitem{Wang2022SchwingerBosonMetallicSpinLiquid} J. Wang and Y.-F. Yang, 
$\mathbb{Z}_2$ metallic spin liquid on a frustrated Kondo lattice,
Phys. Rev. B {\bf 106}, 115135 (2022).

\bibitem{Pairault1998StrongCoupling} S. Pairault, D. S\'{e}n\'{e}chal, and A.-M. S. Tremblay, 
Strong-Coupling Expansion for the Hubbard Model,
Phys. Rev. Lett. {\bf 80}, 5389 (1998).

\bibitem{Pairault2000StrongCoupling} S. Pairault, D. S\'{e}n\'{e}chal, and A.-M. S. Tremblay, 
Strong-coupling perturbation theory of the Hubbard model,
Eur. Phys. J. B {\bf 16}, 85 (2000).

\bibitem{Rubtsov2008DualFermion}  A. N. Rubtsov, M. I. Katsnelson, and A. I. Lichtenstein,
Dual fermion approach to nonlocal correlations in the Hubbard model,
Phys. Rev. B {\bf 77}, 033101 (2008).

\bibitem{Brener2008DualFermion}  S. Brener, H. Hafermann, A. N. Rubtsov, M. I. Katsnelson, and A. I. Lichtenstein,
Dual fermion approach to susceptibility of correlated lattice fermions,
Phys. Rev. B {\bf 77}, 195105 (2008).

\bibitem{Rubtsov2009DualFermion}  A. N. Rubtsov, M. I. Katsnelson, A. I. Lichtenstein, and A. Georges,
Dual fermion approach to the two-dimensional Hubbard model: Antiferromagnetic fluctuations and Fermi arcs,
Phys. Rev. B {\bf 79}, 045133 (2009).

\bibitem{Li2015DualFermion}  G. Li,
Hidden physics in the dual-fermion approach: A special case of a nonlocal expansion scheme,
Phys. Rev. B {\bf 91}, 165134 (2015).

\bibitem{HewsonBook}  A. C. Hewson, 
\textit{The Kondo Problem to Heavy Fermions} 
(Cambridge University Press, Cambridge, England, 1997).

\bibitem{ColemanBook}  P. Coleman, 
\textit{Introduction to Many-Body Physics} 
(Cambridge
University Press, Cambridge, England, 2015).

\bibitem{Okamoto2005DMFT_SquareLattice} S. Okamoto, A. Fuhrmann, A. Comanac, and A. J. Millis,
Benchmarkings for a semiclassical impurity solver for dynamical-mean-field theory: Self-energies and magnetic transitions of the single-orbital Hubbard model,
Phys. Rev. B {\bf 71}, 235113 (2005).

\bibitem{Kusunose2006DMFT_SquareLattice} H. Kusunose,
Influence of Spatial Correlations in Strongly Correlated Electron Systems: Extension to Dynamical Mean Field Approximation,
J. Phys. Soc. Jpn. {\bf 75}, 054713 (2006).

\bibitem{Fuhrmann2007DMFT_SquareLattice} A. Fuhrmann, S. Okamoto, H. Monien, and A. J. Millis,
Fictive-impurity approach to dynamical mean-field theory: A strong-coupling investigation,
Phys. Rev. B {\bf 75}, 205118 (2007).

\bibitem{Zhang2007DMFT_SquareLattice}  Y. Z. Zhang and M. Imada,
Pseudogap and Mott transition studied by cellular dynamical mean-field theory,
Phys. Rev. B {\bf 76}, 045108 (2007).

\bibitem{Wang2009DMFT_SquareLattice}X. Wang, E. Gull, L. de' Medici, M. Capone, and A. J. Millis,
Antiferromagnetism and the gap of a Mott insulator: Results from analytic continuation of the self-energy,
Phys. Rev. B {\bf 80}, 045101 (2009).

\bibitem{Zitko2009DMFT_SquareLattice} R. \v{Z}itko, J. Bon\v{c}a, and T. Pruschke,
Van Hove singularities in the paramagnetic phase of the Hubbard model: DMFT study,
Phys. Rev. B {\bf 80}, 245112 (2009).

\bibitem{Ribic2018DMFT_SquareLattice} T. Ribic, P. Gunacker, and K. Held,
Impact of self-consistency in dual fermion calculations,
Phys. Rev. B {\bf 98}, 125106 (2018).

\bibitem{Park2008CDMFT}  H. Park, K. Haule, and G. Kotliar,
Cluster Dynamical Mean Field Theory of the Mott Transition,
Phys. Rev. Lett. {\bf 101}, 186403 (2008).

\bibitem{Mazitov2022DMFTChargeSusceptibility}  T. B. Mazitov and A. A. Katanin, 
Local magnetic moment formation and Kondo screening in the half-filled single-band Hubbard model,
Phys. Rev. B {\bf 105}, L081111 (2022).

\bibitem{Tan2022MottCriticalityReview}  Y. Tan, V. Dobrosavljevi\'{c}, and L. Rademaker, 
How to Recognize the Universal Aspects of Mott Criticality?
Crystals {\bf 12}, 932 (2022).

\bibitem{Pustogow2018MottPhaseDiagram}  A. Pustogow, M. Bories, A. L\"{o}hle, R. R\"{o}sslhuber, E. Zhukova, B. Gorshunov, S. Tomi\'{c}, J. A. Schlueter, R. H\"{u}bner, T. Hiramatsu, Y. Yoshida, G. Saito, R. Kato, T.-H. Lee, V. Dobrosavljevi\'{c}, S. Fratini, and M. Dressel,
Quantum spin liquids unveil the genuine Mott state,
Nat. Mater. {\bf 17}, 773 (2018).

\bibitem{Arovas1988SchwingerBoson} D. P. Arovas and A. Auerbach,
Functional integral theories of low-dimensional quantum Heisenberg models,
Phys. Rev. B {\bf 38}, 316 (1988).

\bibitem{Read1991SchwingerBoson} N. Read and S. Sachdev,
Large-$N$ expansion for frustrated quantum antiferromagnets,
Phys. Rev. Lett. {\bf 66}, 1773 (1991).

\bibitem{Sachdev1992SchwingerBoson} S. Sachdev,
Kagom\'{e}- and triangular-lattice Heisenberg antiferromagnets: Ordering from quantum fluctuations and quantum-disordered ground states with unconfined bosonic spinons,
Phys. Rev. B {\bf 45}, 12377 (1992).

\bibitem{Terashige2019DHbindingExperiment}  T. Terashige, T. Ono, T. Miyamoto, T. Morimoto, H. Yamakawa, N. Kida, T. Ito, T. Sasagawa, T. Tohyama, and H. Okamoto, 
Doublon-Holon Pairing Mechanism via Exchange Interaction in Two-Dimensional Cuprate Mott Insulators, 
Sci. Adv. {\bf 5}, eaav2187 (2019).

\bibitem{Schmitt-Rink1988SpinPolaron}  S. Schmitt-Rink, C. M. Varma, and A. E. Ruckenstein, 
Spectral Function of Holes in a Quantum Antiferromagnet,
Phys. Rev. Lett. {\bf 60}, 2793 (1988).

\bibitem{Kane1988SpinPolaron}  C. L. Kane, P. A. Lee, and N. Read, 
Motion of a single hole in a quantum antiferromagnet,
Phys. Rev. B {\bf 39}, 6880 (1989).

\bibitem{Martinez1988SpinPolaron} G. Martinez and P. Horsch, 
Spin polarons in the $t$-$J$ model,
Phys. Rev. B {\bf 44}, 317 (1991).

\bibitem{Raimondi1993GaussianofKR} R. Raimondi and C. Castellani,
Lower and upper Hubbard bands: A slave-boson treatment,
Phys. Rev. B {\bf 48}, 11453(R) (1993).

\bibitem{Mazitov2022DMFTResistivity}  T. B. Mazitov and A. A. Katanin,
The effect of local magnetic moments on spectral properties and resistivity near the interaction- and doping induced Mott transitions,
arXiv:2207.14065 (2022).

\end{thebibliography}
\end{document}